\documentclass[aps,pr,preprint,groupedaddress]{revtex4-2}
\input epsf.sty
\usepackage{graphicx,amsmath,amssymb,xcolor,hyperref,ulem}
\hypersetup{colorlinks=true,linkcolor=blue}

\usepackage{color}
\definecolor{red}{cmyk}{0,1,0.50002,0}
\definecolor{blue}{rgb}{0,0.49995,1}
\definecolor{green}{cmyk}{1,0,0.49998,0}
 
\usepackage{graphics,amsmath,amssymb}

\def\lromn#1{\uppercase\expandafter{\romannumeral#1}}

\begin{document}
\begin{flushright}
\today \\
\end{flushright}

\begin{center}
\begin{large}
\textbf{
Observational tests for a class of scalar-tensor gravity
}
\end{large}
\end{center}

\author{Kunio Kaneta}
\email{kaneta@ed.niigata-u.ac.jp}
\affiliation{
Faculty of Education, Niigata University, \\
Niigata 050-2181, Japan}

\author{Kin-ya Oda}
\email{odakin@lab.twcu.ac.jp}
\affiliation{Department of Mathematics, 
Tokyo Woman's Christian University, \\
Tokyo 167-8585, Japan}

\author{Motohiko Yoshimura}
\email{yoshim@okayama-u.ac.jp}
\affiliation{Research Institute for Interdisciplinary Science,
Okayama University \\
Tsushima-naka 3-1-1 Kita-ku Okayama
700-8530 Japan}

\date{\today}

\vspace{2cm}
\begin{abstract}
We study observational bounds in  a class of scalar-tensor gravity theories
recently proposed.
Either an upper or lower bound on a conformal factor in these theories
 is derived from null  observation in composition dependent fifth force search,
microscope mission.
The important case of a lower bound implies that future improved
observations have chances of verifying this class of theories.
Future prospect for a particular type of observation is mentioned.
The considered class of scalar-tensor gravity was shown elsewhere
to explain the conversion of inflationary early phase to late time
quintessence type dark energy.

\end{abstract}


\maketitle

\newpage

\begin{enumerate} \item {\bf Introduction}

Since the advent of inflationary theory \cite{inflation},
\cite{inflation 2}, \cite{cosmology textbooks},
a scalar field, usually called inflaton,
is of central importance in cosmology.
Another twist has been added by the discovery of dark energy \cite{dark energy},
which may be described by a quintessence field \cite{quintessence}.
How these two accelerating universes may or may not be related is, in our view,
a hint towards deeper understanding of our universe, or even a hint
towards a modified gravity beyond general relativity.

We have recently developed how two accelerating universes are related 
by a common origin \cite{koy 24-2}, \cite{koy 25-1}.
Our theoretical model generalizes the original Jordan-Brans-Dicke gravity \cite{jbd} and
its massive extension  \cite{ejbd}.
It was constructed solely by the principle of gauge invariance of standard particle physics,
or its grand unified extension, its five independent
pieces of gauge invariant terms being treated separately with different conformal factors. 
We adopt here a simplified version of this class of extended Jordan-Brans-Dicke (eJBD) gravity,
as elaborated in \cite{koy 25-1}.

We shall briefly recapitulate main features of these models to the extent relevant to our
discussions below.
In the simplified eJBD gravity the relevant part of lagrangian density in the Einstein metric frame
is given by
\begin{eqnarray}
&&
{\cal L} = - \frac{M_{\rm P}^2}{2} R
+ \frac{1}{2} (\partial \chi)^2 - V_{\chi} - \frac{1}{4} F_{\mu\nu}F^{\mu\nu} 
- \bar{\psi}  \left(  \gamma   ( i \partial - e^{-\gamma_g \chi/M_{\rm P} } g A) 
- e^{- \gamma_y \chi/M_{\rm P} } y_{\psi} H  \right) \psi
+ \cdots
\,,
\label {standard lagrangian}
\\ &&
V_{\chi} = V_0 \, (\frac{\chi}{M_{\rm P}})^2 \,e^{ - \gamma_{\chi} \chi/M_{\rm P}}
\,, \hspace{0.3cm}
V_0 > 0
\,, \hspace{0.3cm}
\gamma_{i} > 0 
\,, \hspace{0.3cm}
i = \chi\,, g\,, y
\,,
\label {ejbd field potential}
\end{eqnarray}
with $M_{\rm P} = 1/\sqrt{8\pi G} \sim 2.4 \times 10^{18}\,$GeV.
We omitted gauge multiplet structures, hence 
relevant matrices
 of non-Abelian theories are given only symbolically.
We restricted, in this formula,  conformal coupling factors $\gamma_i$ to 
those of standard model of particle physics, 
but mention later its extension to grand unified theories.

One can achieve a successful inflation for parameter choices, 
$\gamma_{\chi} \sim 0.1\,, V_0 = (O(0.5 \sim 1)\times 10^{16}
{\rm GeV} )^4$, as shown in \cite{koy 23}.
The potential $V_{\chi}$ has a maximum at $\chi= 2 M_{\rm P}/\gamma_{\chi}$,
and eJBD $\chi$ field is assumed to be in the left to this maximum, 
$\chi < 2 M_{\rm P}/\gamma_{\chi}$ initially.
Inflation is realized by the slow roll towards the potential minimum at the field origin $\chi=0$.

Right after inflation the gauge symmetry of particle physics is spontaneously broken,
since there is no temperature correction to the potential in the empty universe.
eJBD scalar field $\chi$ couples with the trace of energy-momentum tensor, and the trace
at the tree level of perturbation theory
consists of contributions of massive particles with mass parameters appearing explicitly.
There is additional contribution at high energy scale after inflation 
due to the trace anomaly,
and this contribution may help since masses of standard model are too small.

Furthermore, in grand unified theories there are several heavy particles;
considering easier inflaton decay due to favorable mass relation, we
focus on three Majorana leptons $N_R = (1+\gamma_5)N/2$ 
(equivalent to right-handed 2-spinor) some of which is assumed lighter than the half
of inflaton mass.
The major inflaton coupling to ordinary particles is then
\begin{eqnarray}
&&
\gamma_J \, \frac{ \chi}{M_{\rm P}}  \sum_A\,
\frac{b_A \alpha_A^2}{8\pi} \, (F^{ A\, (1)} )^{\rho\sigma} \,F^{ A\, (2)} _{\rho\sigma} 
+  \, \gamma_N\, \frac{ \chi}{M_{\rm P}}\,  M_R (\overline{N^C}  N_R + ({\rm h.c.})\, )
\,.
\label {trace anomaly of gauge-pair}
\end{eqnarray}
 $\alpha_A = g_A^2/4\pi$ are various gauge couplings, and $b_A$ is the coefficient of
renormalization group. 
The heavy Majorana mass term $\propto M_R $ violates the lepton number
with $C$ denoting the charge conjugation.
This inflaton coupling is  more general than given in \cite{koy 25-1}.

The conformal factor $\gamma_J$ is related to the metric change in the Einstein frame
of our model, and there exists a severe bound $\gamma_J < 3 \times 10^{-3}$ from
solar system tests on PPN parameters \cite{gr tests}.

Our scalar-tensor theory of gravity does not belong to a
much simplified class of theories often taken up in the literature \cite{gr tests}, \cite{will},
\cite{def}.

$\chi$ field decay into   Majorana particle pairs
given by (\ref{trace anomaly of gauge-pair})  occurs frequently, and this necessarily
induces instability of tachyon type potential correction, quite opposite to
the finite temperature mass correction.
As shown elsewhere \cite{koy 25-1}, this provides a mechanism of
inflaton conversion into quintessence field.
The instability does not continue forever, and is ended during an epoch
towards thermalized hot big bang universe.

It is of crucial importance how eJBD scalar field $\chi$ couples to matter fields, in order to
verify predictions of the theory at recent epochs of cosmological evolution.
Three dimensionless conformal parameters $\gamma_g, \gamma_y, \gamma_N$ 
in the equation (\ref{standard lagrangian}) and (\ref{trace anomaly of gauge-pair})
characterize the strength of inflaton coupling to matter.
One of them $\gamma_N$ may control preheating and thermalization right after inflation,
but it is irrelevant to cosmology after nucleo-synthesis, the main theme of the present work.

Throughout the present work we use the unit of $\hbar = c =1$.

\vspace{0.5cm} 
\item {\bf Fundamental issues related to electromagnetic gauge invariance}

We need to clarify what the conformal factor $e^{-\gamma_g \chi/M_{\rm P}}$
in the lagrangian (\ref{ejbd field potential}) implies since electromagnetic
interaction is not written in terms of the covariant derivative.
The U(1)$_{\rm EM}$ gauge transformation in our eJBD model is modified, and
the correct gauge transformation is
\begin{eqnarray}
&&
A_{\mu} \rightarrow A_{\mu} - \partial_{\mu} (g_{\chi}^{-1} \Lambda) 
\,, \hspace{0.3cm}
\psi \rightarrow e^{i \Lambda' } \psi
\,, \hspace{0.3cm}
g_{\chi} = g e^{- \gamma_{\chi} \chi/M_{\rm P} }
\,,
\\ &&
\partial_{\mu} \Lambda' =\partial_{\mu} \Lambda
+  \gamma_{\chi} \frac{\partial_{\mu} \chi }{M_{\rm P} } \Lambda
\,.
\end{eqnarray}
Accordingly, the charged current $j_{\mu} = \bar{\psi} \gamma_{\mu} \psi$
is not conserved, $\partial^{\mu} j_{\mu} = \gamma_g\, j_0\, \dot{\chi}/M_{\rm P} \neq 0 $,
while the electromagnetic U(1)$_{\rm EM}$ gauge invariance is maintained by
introducing the extra term $\propto \dot{\chi}$ in the fermion
$\Lambda'$ phase transformation.
One cannot gauge away the conformal factor $e^{-\gamma_g \chi/M_{\rm P}}$
because there is no freedom left
for rescaling fields to change the equal-time canonical anti-commutation relation 
between $\psi$ and $\psi^{\dagger}$.

The action principle of our system leads to the Maxwell equation in cosmic medium,
\begin{eqnarray}
&&
\partial_{\mu}F^{\mu \nu} = e \exp[-\gamma_{g} \frac{\chi(t)}{M_{\rm P}} ]\, j^{\nu}
\,, \hspace{0.5cm}
 j^{\nu} = \sum_f q_f \bar{f}\gamma^{\nu} f
\,.
\end{eqnarray}
A part of Maxwell equation, the Poisson equation with $\vec{E} = \vec{\nabla} \phi$,
is then
\begin{eqnarray}
&&
\vec{\nabla}^2 \phi = - e \exp[-\gamma_{g} \frac{\chi(t) }{M_{\rm P}} ]\, j_0
\,.
\end{eqnarray}
From electrostatics one concludes that 
our model predicts the
cosmic medium endowed by a time varying dielectric constant of
$\epsilon= \exp[\gamma_{g} \frac{\chi(t)}{M_{\rm P}} ] $.
The speed of light is thus changed from that in vacuum by an amount,
\begin{eqnarray}
&&
v= \frac{1}{\sqrt{\epsilon}} = \exp[ - \frac{1}{2} \gamma_{g} \frac{\chi(t) - \chi(t_0)}{M_{\rm P}} ] > 1
\,,
\end{eqnarray}
the last inequality being due to the decreasing eJBD field behavior, 
$\chi(t) < \chi(t_0)$ for $t < t_0$ (the present time).
If one interprets this as an effective photon mass, the photon behaves as a tachyon.

Observed simultaneous arrival of
 gravitational wave and electromagnetic radiation in NS-NS merger event
\cite{gw ns-ns merger}
implies that $v-1 $ is severely bounded by $O(10^{-16})$.
This means that $\gamma_g \dot{\chi}(t_0) (t_0 - t) /2M_{\rm P} < O(10^{-16})$ at recent epochs
of cosmic evolution.
This would imply either an extremely  stringent bound on $\gamma_g$ or
the vanishing $\gamma_g$ as a likely possibility.

\vspace{0.5cm} 
\item {\bf Constraints from time variation of fine structure constant}

The fine structure constant variation at times $t< t_0$ is described by the formula,
\begin{eqnarray}
&&
\frac{\delta_{t} \, \alpha_{\rm eff} }{\alpha} = \left(
1 - \exp[- \gamma_g \frac{\chi(t) - \chi(t_0)}{M_{\rm P}} ]
\right)
e^{- \gamma_y \chi(t_0)/M_{\rm P}} 
\,.
\label {effective alpha}
\end{eqnarray}
Note that our model predicts time variation in a definite way as a function
of inflaton field $\chi(t)$.
Many other JBD theories do not provide explicitly time varying function of scalar field,
and there is often no unambiguous prediction of fine structure variation
in these cases.

\begin{figure*}[htbp]
 \begin{center}
\includegraphics[width=0.6\textwidth]{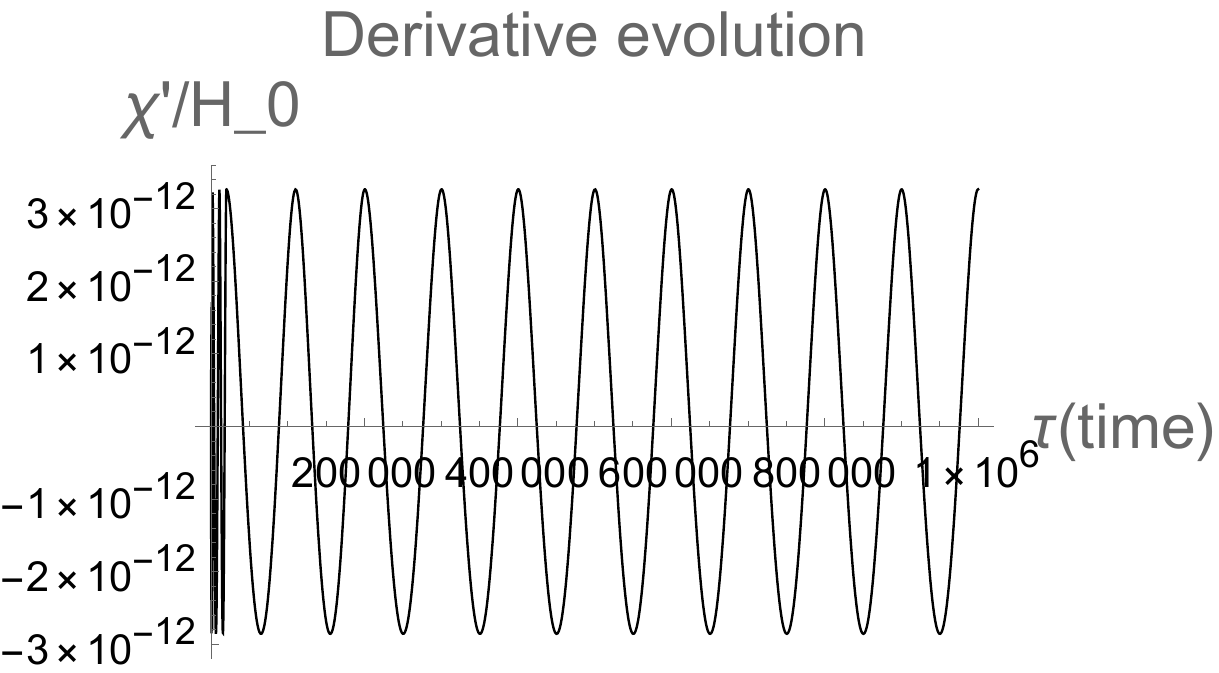}
   \caption{
Time variation of field derivative divided by the Hubble rate: 
numerical solution of (\ref{late time field evol eq}).
Field value remains to be a constant $\sim 500$ over this time range of $\leq 10^{11}$ years. 
}
   \label {field derivative}
 \end{center} 
\end{figure*}

The factors in RHS of (\ref{effective alpha})  monotonically increase,
hence one derives upper bound of conformal coupling. 
Numerical solutions of eJBD field equation at recent epochs, as shown in Fig(\ref{field derivative}),
suggest 
\begin{eqnarray}
&&
\frac{1}{H_0}
\frac{d \chi}{ d t} = O(10^{-12} \sim 10^{-11} )
\,,
\end{eqnarray}
with $H_0$ the present Hubble constant,
hence 
\begin{eqnarray}
&&
\delta_t\, \alpha = - 1.3 \times O( 10^{-21} 
\sim 10^{-20} ) \, \gamma_g  \, \frac{\Delta t}{ 1 {\rm year}}
\,.
\end{eqnarray}
The amplitude of field derivative is of order $3 \times 10^{-12}$
in the Hubble unit, superimposed oscillation. 
Time duration of 1 year gives corresponding field derivative  
$\sim 4 \times 10^{-12}$ in the Hubble unit.

There are a variety of different observational tests conducted on time variation of fine structure constant in the past \cite{gr tests}.
Laboratory-type experiment of atomic clock
\cite{atomic clock: alpha} has been conducted over
several months at the present time of cosmic evolution, while
the Oklo null-result  \cite{oklo} compares the present value of
fine structure constant  to that at a past nuclear fission event  that
occurred 18 billion years ago.
The bound of quasar absorption spectra uses fine structure splitting of alkali-like atoms 
\cite{quasar bound} at a redshift of a few.

Two atomic observations \cite{atomic clock: alpha}, \cite{quasar bound}
use electromagnetic transitions between energy levels, energies being
functions of $\alpha^n m_e$ with different powers $n$ depending on the nature
of radiative transitions.
In our model both $\alpha$ and the electron mass $m_e$ may time vary,
which has to be considered properly.
The situation in the Oklo bound is far more complicated:
the bound is sensitive to neutron capture cross section
of a process Sm$^{149} (n, \gamma)$Sm$^{150}$ proceeding via
resonance formation.
The small resonance energy of order 100 meV arises from
a nearly complete cancellation of Coulomb and strong interactions
of protons inside Sm nuclei.
To make derivation of the bound meaningful,
the authors of \cite{oklo} assume constancy of mass parameters like the proton mass.

\begin{figure*}[htbp]
 \begin{center}
\includegraphics[width=0.6\textwidth]{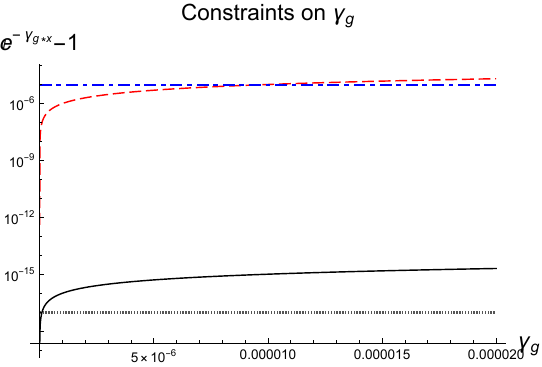}
   \caption{
Constraints on $\gamma_g$ using the bound on $\gamma_y$;
$\gamma_g < O(1 \times10^{-5} )$  obtained from null observation of fine structure
constant variation in distant quasars shown in two colors.
A laboratory bound of atomic clock \cite{atomic clock: alpha} 
shown in blacks gives a severer constraint 
$\gamma_g < O(1 \times 10^{-7} )$.
}
   \label {oklo and clock bound}
 \end{center} 
\end{figure*}

Thus, laboratory type experiments (for example, \cite{atomic clock: alpha} giving $< O(10^{-17})$ 
bound)  give a constraint of order, $\gamma_g < O(1 \times  10^{-7} )$.
The null-measurement at high redshift observations \cite{quasar bound},
gives a bound $\gamma_g < O( 1 \times 10^{-5}) $.
The Oklo bound \cite{oklo} is not included in our analysis,
due to ambiguities of interpretation \cite{alpha variation 2}.

We conclude that the recent constraint on $\gamma_g$ from simultaneous measurement
of gravitational and electromagnetic waves from NS-NS merger event
is much more stringent than bounds from the fine structure constant variation.
A likely possibility is the vanishing $\gamma_g=0$.

We turn to another conformal factor $\gamma_y$ in the next section.

\vspace{0.5cm} 
\item {\bf Scalar exchange force at recent epochs of cosmic evolution}

In the present work we focus on recent epochs of the redshift factor $z+1 \leq O({\rm a\; few})$.
eJBD field mass is of the Horizon mass $O(10^{-32})$eV, and we may regard the field effectively massless
in events we consider.
Field coupling determined by (\ref{standard lagrangian}) however differs from the original JBD theory.
Let us work out the potential due to exchange of field quantum between two particle.
Field quantum defined by $\delta \chi (x)= \chi(x) - \chi_0(t)$ as deviation from
the universal $\chi_0(t)$ gives rise to a variety of linear terms in $\delta \chi$ in the lagrangian.
In the spontaneously broken phase we first focus on coupling via fermion mass terms, 
\begin{eqnarray}
&&
\delta {\cal L}_{\chi f} = - \frac{\delta \chi} {M_{\rm P}} \, \gamma_y\, e^{ - \gamma_y x_t}  \sum_f  m_f\, \bar{f} f 
\,, \hspace{0.5cm}
x_t = \frac{\chi_0(t)}{M_{\rm P}}
\,.
\label {yukawa and fsc}
\end{eqnarray}
$x_t $ refers its value at a cosmic time $t$ with the present time $t=t_0$.

Quantum $\delta \chi$ exchange potential between two fundamental fermions,
quarks and leptons,  at rest  is given by
\begin{eqnarray}
&&
V_{12} = - \frac{ 2 G}{r } (\gamma_y\, e^{- \gamma_y x_0})^2 m_1 m_2 
\,,
\end{eqnarray}
with $G$ the gravitational constant.
Hence field exchange force is weaker than gravitational force by
the amount that contains $\gamma_y$ dependent factors.

The exchange potential between two nucleons and neutral atoms is more complicated
due to the quark condensate present  both in vacuum \cite{svz}
and in nuclear medium \cite{condensates in nuclear matter}.
The conclusion of past works is 
\begin{eqnarray}
&&
\int_{\rm nucleon} d^3x \, \sum_{i=u,d} m_i 
\langle \bar{q_i} q_i \rangle_{\rm nucleon} \sim - \left( 94 \, {\rm MeV} + 8.6 (m_u+m_d)
\right) 
\,.
\end{eqnarray}
As a representative value we shall take  $- 1/7$ of nucleon mass.
Between neutral atoms there are extra contributions of the mass of $Z$ electrons. 
Hence we find $\chi$ quantum exchange gives potentials,
\begin{eqnarray}
&&
V(N_1, N_2) = 
- \frac{2G}{ r}
\gamma_y^2 e^{- 2 \gamma_y \, \chi_0/M_{\rm P}}\,
( - 130 {\rm MeV} )^2 \approx  - \frac{2G}{ r} 
\gamma_y^2 e^{- 2 \gamma_y \, \chi_0/M_{\rm P}}\, \frac{m_N^2}{49}   
\,,
\\ &&
V( ^{Z_1}\!A_1-^{Z_2}\!A_2) \approx
- \frac{2G}{ r} \gamma_y^2  e^{- 2 \gamma_y \chi_0/M_{\rm P} }
\,  ( - \frac{m_N}{7} \, A + m_e Z )_1 \, ( -  \frac{m_N}{7}  \, A + m_e Z )_2
\,,
\end{eqnarray}
for two nucleons of mass $m_N$ and two neutral atoms, respectively.

\vspace{0.5cm} 
\item {\bf Constraint from composition dependent fifth force}

The best observational limit  of the  composition dependent weak equivalence principle is
provided by MICROSCOPE mission, which gives the limit 
on E$\ddot{{\rm o}}$tv$\ddot{{\rm o}}$s ratio $\eta ({\rm Pt, Ti}) $
\cite{microscope mission},
\begin{eqnarray}
&&
\eta ({\rm Pt, Ti}) = (-1.5 \pm 2.3 ({\rm stat}) \pm 1.5 ({\rm syst}) )\times 10^{-15}
\,, \hspace{0.3cm} 
\eta(1,2) \equiv \sum_A \eta^A \left( \frac{E_1^A}{m_1} -  \frac{E_2^A}{m_2} \right) 
\,,
\end{eqnarray}
where $E_i^A$ is composition dependent contribution to the inertial mass,
while $m_i$ is the gravitational mass.

The quantum number of material is as follows;
 $Z/A \sim 0.3998 $ for Pt, and $\sim 0.4596 $ for Ti.
A common contribution $8\pi \gamma_y^2 (-0.138)^2$ cancels
in $\eta  ({\rm Pt, Ti})$.
Hence the linear term $\propto Z/A$ remains as the major contribution in our model, to give
\begin{eqnarray}
&&
\eta ({\rm Pt, Ti}) = - 0.138\, (0.4596-0.3998)\frac{m_e}{m_N}  \, 8\pi \,
\gamma_y^2 e^{- 2 \gamma_y \chi_0/M_{\rm P} }
\sim - 1.1 \times 10^{-4}\, \gamma_y^2\, e^{- 2 \gamma_y \chi_0/M_{\rm P} }
\,.
\label {pt-ti firth force}
\end{eqnarray}

It becomes important to relate the quantity 
$\gamma_y^2 e^{- 2 \gamma_y \chi_0/M_{\rm P}}  $ 
that appears in the right hand side (RHS)
to observable cosmological quantities at recent epochs.
We first verify the potential energy dominance over the kinetic energy 
by solving the field equation including both kinetic and potential energies
of the field (assuming the dark energy dominance over the dark matter contribution.
Thus, the evolution equation reads as, at recent epochs using
the dimensionless time $\tau = \sqrt{V_0/M_{\rm P}^2}\, t$ and the dimensionless field
 $x=\chi/M_{\rm P}$,
\begin{eqnarray}
&&
\ddot{x} + \sqrt{3}\, H  \dot{x} +  x ( 2 -\gamma_{\chi} x)e^{-\gamma_{\chi}x} =0
\,, \hspace{0.5cm}
\sqrt{3}\, H = \sqrt{\left( \frac{1}{2}(\frac{dx}{d\tau})^2 + x^2 e^{-\gamma_{\chi} x} \right) }
\,.
\label {late time field evol eq}
\end{eqnarray}
Solutions of this equation with $\gamma_{\chi} = 0.1$ have vindicated that
kinetic energy contributions are much smaller than the potential energy contribution.
Hence, the potential energy dominance is justified.

\begin{figure*}[htbp]
 \begin{center}
\includegraphics[width=0.6\textwidth]{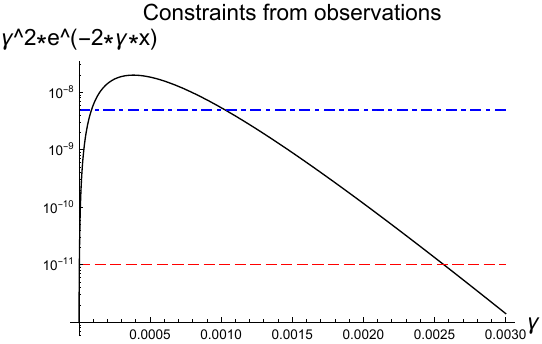}
   \caption{
Observational bounds on the coupling $\gamma_y$ using the theoretical prediction in solid black:
$\gamma_y > O( 0.0026) $ or $\gamma_y < O(3.2 \times 10^{-6} )$  from the limit of
composition dependent fifth force determined from the cross points in dashed orange, and 
$\gamma_y >O( 0.001)$ or, $\gamma_y < O(0.9\times 10^{-4})$, 
from PSR1913+16 cross points in dash-dotted blue.
}
   \label {gammas constraint from observations}
 \end{center} 
\end{figure*}

Assuming the potential energy dominance,
one can relate the needed factor to the Hubble energy $\sim x e^{- \gamma_{\chi} x_0/2}/\sqrt{3}$
with $\gamma_{\chi} \sim 0.1$,
\begin{eqnarray}
&&
V_0\, x_0^2\, e^{-  \gamma_{\chi} x_0}  = 3 (H_0 M_{\rm P})^2 
\,.
\end{eqnarray}
With a range of $V_0 = (0.5 \sim 1) \times 10^{16}$GeV,  $\gamma_{\chi} = 0.1$
and the Hubble energy $H_0 \sim 1.33 \times 10^{-32}$eV,
one derives a value $x_0$ and a relevant function $c_f(\gamma_y)$, 
\begin{eqnarray}
&&
x_0 \sim 2610
\,, \hspace{0.5cm}
c_f(\gamma_y) =
\gamma_y^2 \, e^{- 2 \gamma_y\, x_0} \sim \gamma_y^2\, (1.80 \times 10^{-227} )^{10 \gamma_y}
\,.
\label {gamma constraint eq}
\end{eqnarray}

The left hand side (LHS) quantity $c_f(\gamma_y) $ against $\gamma_y$ has a maximum value 
$\sim 2.0 \times 10^{-8}$ at $\gamma_i \sim 3.8 \times 10^{-4}$.
There may be two markedly different constraints, either an upper bound or a lower bound.
The surprising lower bound  occurs,
because the relevant function $c_f(\gamma_y)$ of $\gamma_y$ has a maximum, and
it necessarily decreases exponentially  towards zero $\gamma_y$ beyond this maximum.
Suggested finite values of $\gamma_y$ in the case of lower bound are very exciting, since
these are subject to a variety of observational tests.

MICROSCOPE mission gives a severe upper bound or
an interesting lower bound on our model parameter $\gamma_y$.
We show in Fig(\ref{gammas constraint from observations}) constraints derived
from MICROSCOPE mission and the other one discussed below.  
The derived lower bound $\gamma_y > O(1/380)$ implies an exciting possibility
for future observation.

\vspace{0.5cm} 
\item {\bf Future prospect: Application to dark wave emission in binary systems}

The first indication of gravitational wave (GW) emission was found in binary pulsar PSR1913+16,
evidenced by orbital parameter variation of two neutron stars of nearly common mass $\sim 1.4\, m_{\rm solar}$
and eccentricity $\sim 0.6$, as summarized in \cite{st textbook}. 
Agreement of data with prediction by general relativity may place constraint on
other sources of energy loss of this binary system.
More recently, advanced LIGO detectors have  observed direct gravitational wave emission
from many binary merger systems, mostly merger of two black holes.
But there are a couple of GW candidates for neutron star (NS) and black hole merger(BH).

Mergers of NS and BH are best suited to test our class of scalar-tensor gravity.
The no-hair conjecture \cite{no-hair conjecture} against the scalar charge of NS
suggests that a complete discharge of neutron star should occur
 prior to merger into a single black hole.
Related to this,
unstable linearized perturbation modes
around spherically symmetric static solution, which has necessarily the curvature
singularity, were found \cite{jbd instability}.
Since  dynamical  gravitational collapse proceeds without
developing unstable modes, this analysis suggests the no-hair conjecture.

How and when the scalar discharge occurs may greatly influence
gravitational wave emission in NS-BH merger, and we encourage
careful observational watch.
Importance of the problem is well recognized in the community, for instance as seen
in \cite{will} which stresses relevance of scalar-tensor theory
in relation to the dark energy problem.
There is also lower order PPN calculations in \cite{dipole emission}.

A report on gravitational wave emission from known 200 pulsars
\cite{scalar-wave emission from pulsars}, without any signal, is not
particularly useful to our problem.
Compact binaries made of pulsar and black hole have better chances
of testing our class of scalar-tensor gravity,
and we look forward to see dedicated observational study for a long duration of time.

\vspace{0.5cm} 
\item {\bf Summary}

We have discovered an exciting possibility of lower bound on a model parameter
in a class of scalar-tensor gravity from null result of composition dependent fifth force;
microscope mission.
The lower bound opens a window of finding deviation from general relativity in
forthcoming observations.
We have hinted as a future prospect
 a novel way of observing a dramatic event that may affect GW emission
at the time of complete scalar discharge.
Observational constraints on the conformal factor of gauge coupling $\gamma_g$
is not as strong as the Yukawa conformal coupling $\gamma_y$.
Moreover, the recent observation of simultaneous arrival of gravitational wave
and electromagnetic radiation from NS-NS merger suggests
 the vanishing $\gamma_g$.
Our class of scalar tensor theory can explain how the inflationary phase at early universe is converted to
the quintessence type of dark energy at recent epochs, as shown elsewhere \cite{koy 25-1}.

\end{enumerate}

\vspace{0.5cm}
\begin{acknowledgments}
We appreciate  S. Morisaki and H. Tagoshi  for a useful comment 
on the solar system bound.
This research was partially
 supported by JSPS KAKENHI  Nos. 21H01107 (KO) and 21K03575(MY).

\end{acknowledgments}

\end{document}